\documentclass[aps,10pt,twocolumn,superscriptaddress]{revtex4-1}
\usepackage[colorlinks=true,allcolors=blue]{hyperref}
\usepackage{amssymb}
\usepackage{amsmath}
\usepackage{graphicx}
\usepackage[caption=false]{subfig}
\usepackage{amsfonts}
\usepackage{xcolor}
\usepackage{epsfig}
\usepackage{color}
\usepackage{bm}
\usepackage{tabularx}
\usepackage{multirow}
\usepackage{mathtools}
\usepackage{xfrac}
\usepackage{blkarray}
\usepackage{bbold}
\usepackage[mathscr]{euscript}
\usepackage{autobreak}
\usepackage{comment}

\newcommand{\vb}{V$_{\rm B}$}
\newcommand{\cn}{C$_{\rm N}$}
\newcommand{\cb}{C$_{\rm B}$}
\newcommand{\on}{O$_{\rm N}$}
\newcommand{\ob}{O$_{\rm B}$}
\newcommand{\cnvb}{C$_{\rm N}$V$_{\rm B}$}
\newcommand{\onvb}{O$_{\rm N}$V$_{\rm B}$}
\newcommand{\vncb}{V$_{\rm N}$C$_{\rm B}$}
\newcommand{\vnob}{V$_{\rm N}$O$_{\rm B}$}
\begin{document}
\title{Quantum Emission from Coupled Spin Pairs in Hexagonal Boron Nitride}

\author{Song Li}
\email{li.song@wigner.hun-ren.hu}
\affiliation{HUN-REN Wigner Research Centre for Physics, P.O.\ Box 49, H-1525 Budapest, Hungary
}

\author{Anton Pershin}
\affiliation{HUN-REN Wigner Research Centre for Physics, P.O.\ Box 49, H-1525 Budapest, Hungary
}
\affiliation{Department of Atomic Physics, Institute of Physics, Budapest University of Technology and Economics,  M\H{u}egyetem rakpart 3., H-1111 Budapest, Hungary}

\author{Adam Gali}
\email{gali.adam@wigner.hun-ren.hu}
\affiliation{HUN-REN Wigner Research Centre for Physics, P.O.\ Box 49, H-1525 Budapest, Hungary
}
\affiliation{Department of Atomic Physics, Institute of Physics, Budapest University of Technology and Economics,  M\H{u}egyetem rakpart 3., H-1111 Budapest, Hungary}
\affiliation{MTA–WFK Lend\"{u}let "Momentum" Semiconductor Nanostructures Research Group, P.O.\ Box 49, H-1525 Budapest, Hungary}

\date{\today}
\begin{abstract}
Optically addressable defect qubits in wide band gap materials are favorable candidates for room temperature quantum information processing. The two-dimensional (2D) hexagonal boron nitride (hBN) is an attractive solid state platform with a great potential for hosting bright quantum emitters with quantum memories with leveraging the potential of 2D materials for realizing scalable preparation of defect qubits. Although, room temperature bright defect qubits have been recently reported in hBN but their microscopic origin, the nature of the optical transition as well as the optically detected magnetic resonance (ODMR) have been remained elusive. Here we connect the variance in the optical spectra, optical lifetimes and spectral stability of quantum emitters to donor-acceptor pairs (DAP) in hBN by means of \textit{ab initio} calculations. We find that DAPs can exhibit ODMR signal for the acceptor counterpart of the defect pair with $S=1/2$ ground state at non-zero magnetic fields depending on the donor partner. The donor-acceptor pair model and its transition mechanisms provide a recipe towards defect qubit identification and performance optimization in hBN for quantum applications.   
\end{abstract}

\maketitle

%
%
\section{Introduction}
Isolated optically active atomic defects in wide band gap materials serve as single photon emitters (SPE) which are key requirements for quantum information technologies~\cite{zhang2020material, wolfowicz2021quantum}. 
Hexagonal boron nitride (hBN) is a layered van der Waals (vdW) material and a favorable SPE host due to its versatility of fabrication and compatibility with lithographic processing~\cite{cassabois2016hexagonal, caldwell2019photonics, sajid2020single, liu2022spin, montblanch2023layered}. The observed SPEs in hBN feature high brightness, room temperature stability, sharp zero-phonon-lines (ZPL) peaks at around 2~eV and short excited state lifetime~\cite{tran2016quantum, tran2016robust, stern2022room, mendelson2021identifying, guo2023coherent, noh2018stark, xu2021creating, kianinia2018all, li2019purification}. Furthermore, the coherent control of single electron spins has been recently realized in hBN~\cite{chejanovsky2021single}, some of them operating at room temperature~\cite{horder2022coherence, stern2024quantum} where the electron spin initialization and readout is based on optical excitation and emission of the defect spins, called optically detected magnetic resonance (ODMR). 


One major challenge is the identification of exact defect structures of the SPEs and single-spin ODMR centers which is a prerequisite for realizing deterministic formation and control. The observed photoluminescence (PL) spectra exhibit various ZPL energies and phonon sideband (PSB) in the spectrum and many of them show similar optical lineshapes~\cite{tran2016robust}. The observed emissions might come from various defects but the similarities in their optical lineshape also imply common defect types in diverse crystalline environment~\cite{li2024exceptionally, plo2024isotope, iwanski2024revealing}.  

An $S = 1/2$ paramagnetic defect with strong hyperfine interaction of two equivalent nitrogen nuclei has been observed by electron paramagnetic resonance (EPR)~\cite{toledo2020identification} and we assigned this signal to the negatively charged \onvb\ defect, i.e., oxygen substituting nitrogen adjacent to boron-vacancy, based on the excellent agreement between the experimental and simulated EPR spectra~\cite{li2022identification}. Noteworthy, the existence of \onvb\ defect has been confirmed by a following annular dark-field scanning transmission electron microscopy (ADF-STEM) experiment~\cite{li2023prolonged}. In addition, carbon and oxygen substitutions are observed simultaneously nearby with the same technique. This is a strong evidence that an extra charge on \onvb\ defect giving rise to the EPR signal could come from the carbon and oxygen substitution of boron (\cb) and nitrogen (\on), respectively, with donor character~\cite{weston2018native}. In other words, \cb\ or \on\ forms a donor-acceptor pair (DAP) with \onvb\ that may be described as \cb$^+$ -- \onvb$^-$ or \on$^+$ -- \onvb$^-$ in the ground state which has $S=1/2$ ground state coming from the spin density around the \onvb$^-$ part of the DAP. In this sense, the common defect type is \onvb\ acceptor defect and the variance in the optical properties comes from the type and location of the donor partner.  

Here we perform comprehensive theoretical calculations for the optical properties of DAP with different separation distance. We find with our model that the donor (\cb\ and \on) indeed provides an electron towards \onvb\ defect which becomes negatively charged. The distance between donor and acceptor can significantly influence the electronic structure. This could be a possible explanation for the ZPL variation observed in experiments. We find that \on\ -- \onvb\ DAP is photostable with quantum yields similar to that of the isolated negatively charged \onvb . On the contrary, the recombination energy of \cb\ -- \onvb\ develops metastable dim states at certain DAP distances which act as non-radiative decay pathways. The spin flipping in these metastable dim states could mix the doublet and quartet multiplets which causes spin-polarization at the $S=1/2$ ground state of the negatively charged \onvb\ when external magnetic field splits these Kramers-doublets. As a consequence, ODMR spectrum~\cite{li2022identification} may be observed for the pair of \cb\ -- \onvb\ in the $S=1/2$ ground state in the presence of a constant magnetic field.

\section{Results}
\label{sec:results}

\begin{figure}[tb]
\includegraphics[width=\columnwidth]{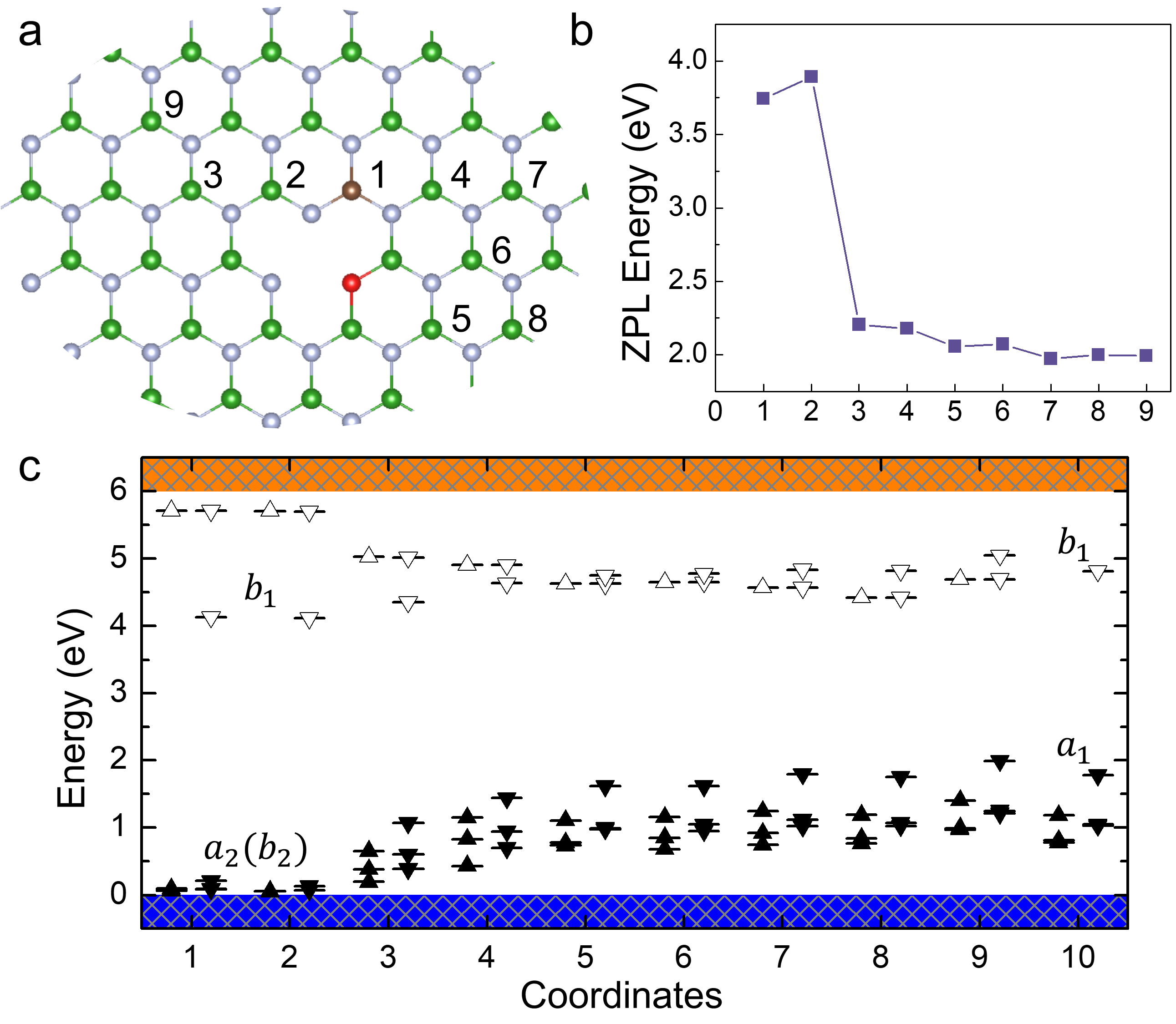}
\caption{\label{Figure1}%
\cb\ -- \onvb\ DAPs in hBN. \textbf{a} The numbers indicate the location of \cb\ at various distances in hBN lattice (boron and nitrogen atoms are gray and green balls, respectively). The \cb\ here is placed to site 1 (brown ball). \textbf{b} The distance-dependent ZPL energies in the intrinsically bright optical transition without considering the non-radiative processes. \textbf{c} The electronic structure of defects in the ground state. Number 10 is the isolated \onvb$^-$ defect for reference.}
\end{figure}

Our modeling relies on recent ADF-STEM results where they were able to image single and multiple carbon and oxygen atoms adjacent to \vb\ monovacancy, in particular, substituting the innermost nitrogen atoms~\cite{li2023prolonged}. However, the monovacancy structures with multiple first-neighbor atom substitutions do not yield bright ZPL emission at 2~eV that has been frequently observed in experiments, as discussed in Supplementary Note 1. Instead, another interesting phenomenon was observed by ADF-STEM where carbon and oxygen substitutional defects exist in vacancy-free regions. The presence of \cb, \cn\ and \on\ defects can be readily confirmed, while \ob\ defect has a high formation energy and therefore is rarely observed~\cite{li2023prolonged}. A previous theoretical study has indicated the donor characteristics of \cb\ and \on\ defects~\cite{weston2018native}. We emphasize that the oxygen related \onvb\ defect may exhibit 2-eV ZPL emission in its negative charge state~\cite{li2022identification}. Hence, we propose that the observed SPE is \onvb$^-$, with the additional charge supplied by either \cb\ or \on. In Supplementary Figure 1, we illustrate that the single electron levels of \cb\ or \on\ are situated at higher energies than the empty state $a_1$ of \onvb\, promoting electron transfer from the donors to \onvb. Furthermore, the total energy of \cb$^+$ -- \onvb$^-$ configuration is lower than that of \cb$^0$ -- \onvb$^0$, confirming the stability of DAP. Noteworthy, the charge transfer within DAP is usually distance-dependent; in wide band gap materials the probability of charge transfer rapidly decays with increasing the distance between the donor and acceptor. Therefore, we focus on several DAP structures characterised by relatively short distances. 

We begin by examining the properties of the \cb\ -- \onvb\ DAP, as illustrated  in Fig.~\ref{Figure1}. The atomic sites or "coordinates" of the donor species with respect to the location of the acceptor species are labeled in Fig.~\ref{Figure1}\textbf{a}. The first-neighbor site, which is capable of forming a direct bond to oxygen, was not considered due to a potential repulsion interaction between \on\ and \cb~\cite{guo2023coherent}. \cb1 and \cb2 sites are close to the oxygen atom, leading to the remarkable influence of the donor on the properties of the DAPs. Following the previously defined notations, as shown in Supplementary Figure 1, the optical transition within \onvb$^-$ occurs between the $a_1$ and $b_1$ levels. However, the strong interaction within \cb1 and \cb2 can change the order of occupied defect levels of \onvb$^-$ and the $a_1$ level shifts down while $a_2$ or $b_2$ level shifts up (the presence of \cb\ makes the symmetry $C_{1h}$ so the original $a_2$ and $b_2$ labels are unified). Since the $a_2$ ($b_2$) orbital has out-of-plane wavefunctions while $b_1$ lies in-plane, the transition dipole moment is along out-of-plane direction with very small magnitude at $\sim$0.12~Debye so the optical transition between them is weak. The bright emission still comes from the $a_1$ to $b_1$ optical transition. The large energy separation between $a_1$ and $b_1$ leads to large ZPL energies. It is 3.7 and 3.9~eV for \cb1 and \cb2, respectively, and locates at ultraviolet region. With increased distance, the energy levels' order restores to the \onvb$^-$ character and the wavefunction of \cb\ retrieves $D_{3h}$ symmetry. Another trend is the unoccupied levels of \cb\ shift downward while the $b_1$ level shifts upward. In \cb5 case, the empty level from \cb\ is below the $b_1$ orbital from \onvb. Nevertheless, the transition dipole moment from $a_1$ to \cb\ is only 0.16~Debye so we focus on the bright emission localized on \onvb\ where the ZPL energies gradually converge to 1.97~eV. 

The above discussed bright emission is a local excitation within \onvb$^-$. Beside this, the charge transfer process from \onvb$^-$ to \cb$^+$ may also luminesce, although it is dim. The dim metastable state (MS) has two sub-states $S = 1/2$ doublet or $S = 3/2$ quartet depending on relative spin orientations within DAP. With Kohn-Sham density functional theory, we could reliably calculate the total energy and geometry of the $S = 3/2$ state. Supplementary Figure 3 shows the Kohn-Sham energy difference for which Kohn-Sham defect states pairs can be associated with the intrinsically dim and bright optical transitions (without considering non-radiative processes and rates here). Generally the bright local excitation is independent on the distance between DAP while the dim one shows a strong relation. The bright emission is the first optical transition in \cb3 and \cb4 and when \cb\ resides at larger distances from \onvb\ then the dim emission has lower energy. This means there could be an observable non-radiative decay path for the local excited state to decay to the ground state. A simple model to capture the dim emission mechanism can be expressed as
\begin{equation}
E(R_i) = E_\text{gap}-(E_\text{D} + E_\text{A}) + \frac{e^2}{\epsilon R_i} \text{,}
\end{equation}
where $E_\text{D}$ and $E_\text{A}$ are the energy levels of the donor and the acceptor in the band gap, respectively. Here, we use the charge transition level (CTL) of \cb($0|+1$) and \onvb($-1|0$) from a previous study~\cite{weston2018native}. $\epsilon$ is the dielectric constant of hBN and $R_i$ is separated distance between donor and acceptor. The last term is the Coulombic interaction and the reciprocal function leads fast convergence of $E(R_i)$ to the CTL difference (1.27~eV) of DAP with large $R_i$, as shown in Supplementary Figure 4. The $S = 3/2$ state can converge properly and it might be a reliable reference to predict the energy of $S = 1/2$ state because the energy difference here is from the spin-spin exchange interaction which is relatively small. The total energy of $S = 3/2$ state lies with 1.68~eV to 2.36~eV higher than that of the ground state depending on the DAP distance and this means that the bright excited state ($\sim$2.0~eV) and the metastable states can switch their energy position.

\begin{figure}
\includegraphics[width=\columnwidth]{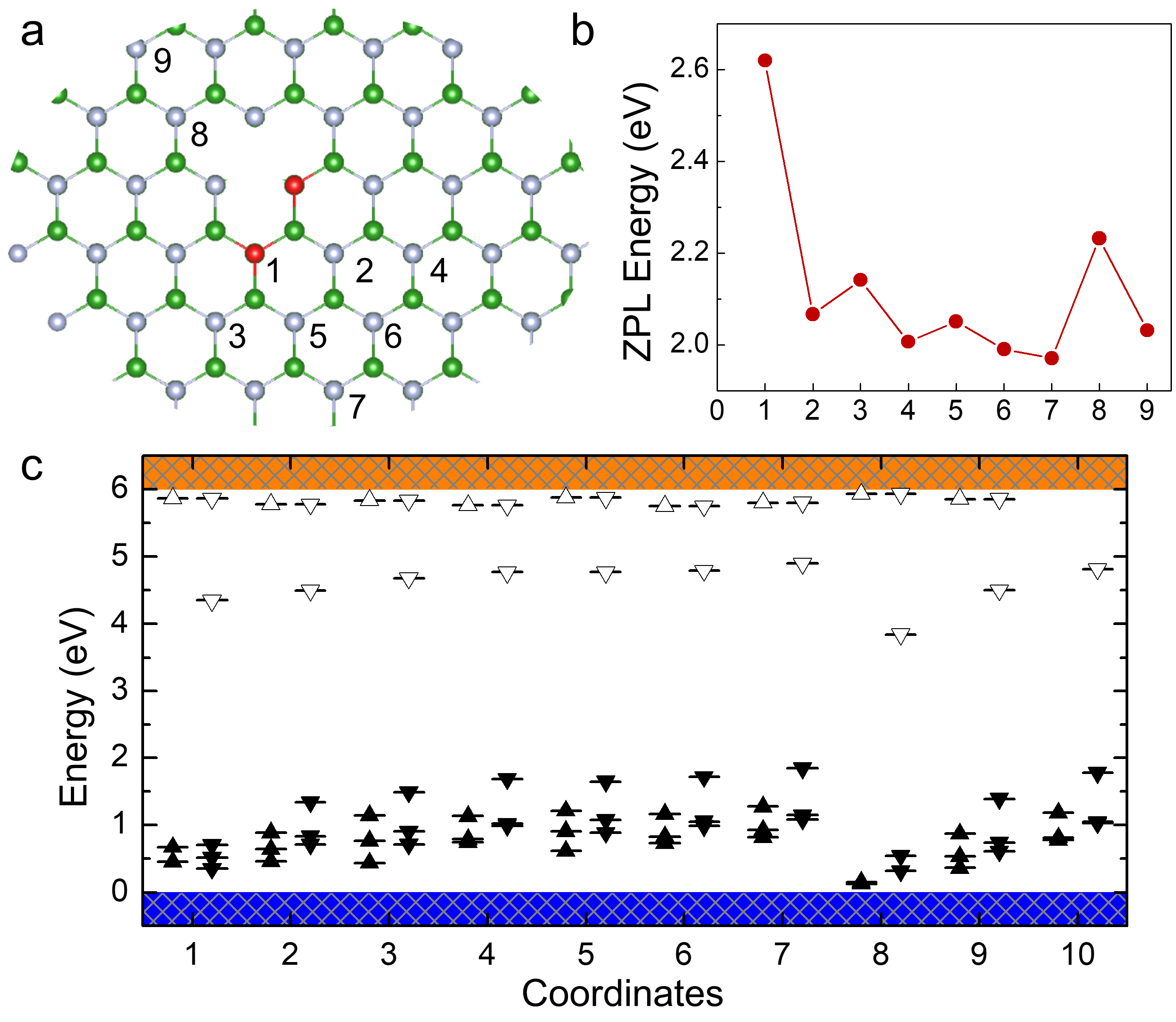}
\caption{\label{Figure2}%
\on\ -- \onvb\ DAPs in hBN. \textbf{a} The numbers indicate the location of \on\ at various distances in hBN lattice (boron and nitrogen atoms are gray and green balls, respectively). The \on\ here is placed to site 1 (red ball). \textbf{b} The distance-dependent ZPL energies. \textbf{c} The electronic structure of defects in the ground state. Number 10 is the isolated \onvb$^-$ defect for reference.}
\end{figure}

\begin{figure}
    \includegraphics[width=\columnwidth]{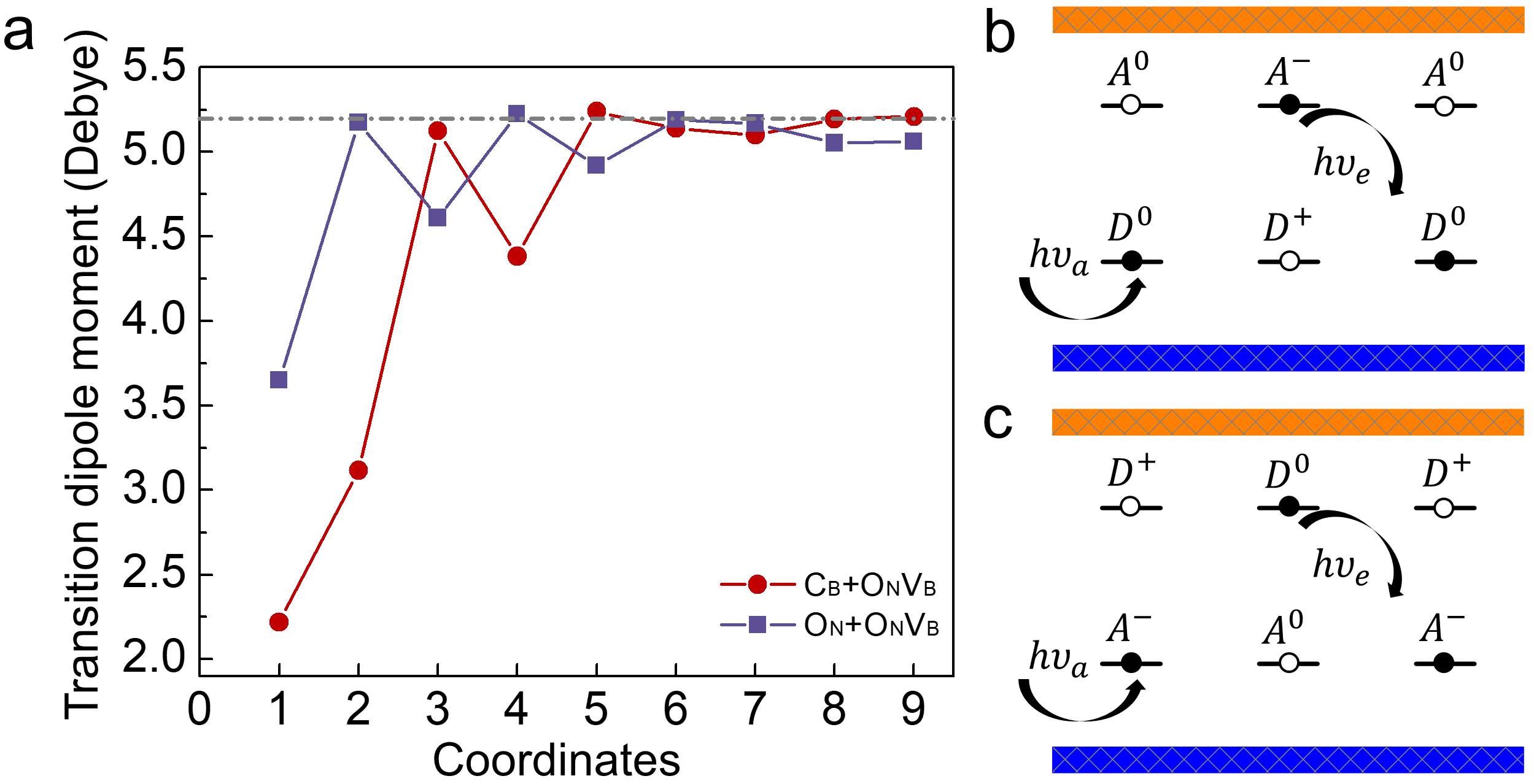}
     \caption{\label{Figure3}%
     Two types of emission from DAPs. (a) The transition dipole moment evolution upon defect configurations we considered is plotted. The dash line indicates the transition dipole moment of \onvb$^-$. The coordinate labels are defined in Figs.~\ref{Figure1} and \ref{Figure2}. (b) Type 1 and (c) type 2 mechanism of optical transition for DAPs in semiconductors. Type 2 may lead to bright emission (see text for details).}
\end{figure}

\on\ -- \onvb\ DAPs are different from \cb\ -- \onvb\ DAPs (see Fig.~\ref{Figure2}). Even with short distance between \on\ and \onvb, the \onvb$^-$ character is still well kept, and this might be due to shallow donor property of \on. The empty levels from \on\ are very close to the conduction band minimum so they have negligible influence on the optical transition within \onvb. Except the \on1 site, the ZPL energies are all below 2.23~eV. Fig.~\ref{Figure3}\textbf{a} shows the transition dipole moment evolution upon the configurations considered. It converges fast to that of the isolated \onvb$^-$ with slightly increased distance between DAP. The large energy difference between \on($0|+1$) and \onvb($-1|0$) results in $E(R_i)$ to be higher than 2.89~eV, e.g., the $S = 3/2$ level of \on4 lies at 3.59~eV above the ground state. Correspondingly, no metastable states arise for \on$^+$ -- \onvb$^-$, which makes the optically and spin active \onvb$^-$ counterpart of DAP a photostable emitter, albeit with a little variation in the ZPL wavelengths depending on the actual distance between the acceptor and the donor. 

\section{Discussion}
\label{sec:discussion}
\begin{figure*}[htbp!]
    \includegraphics[width=2\columnwidth]{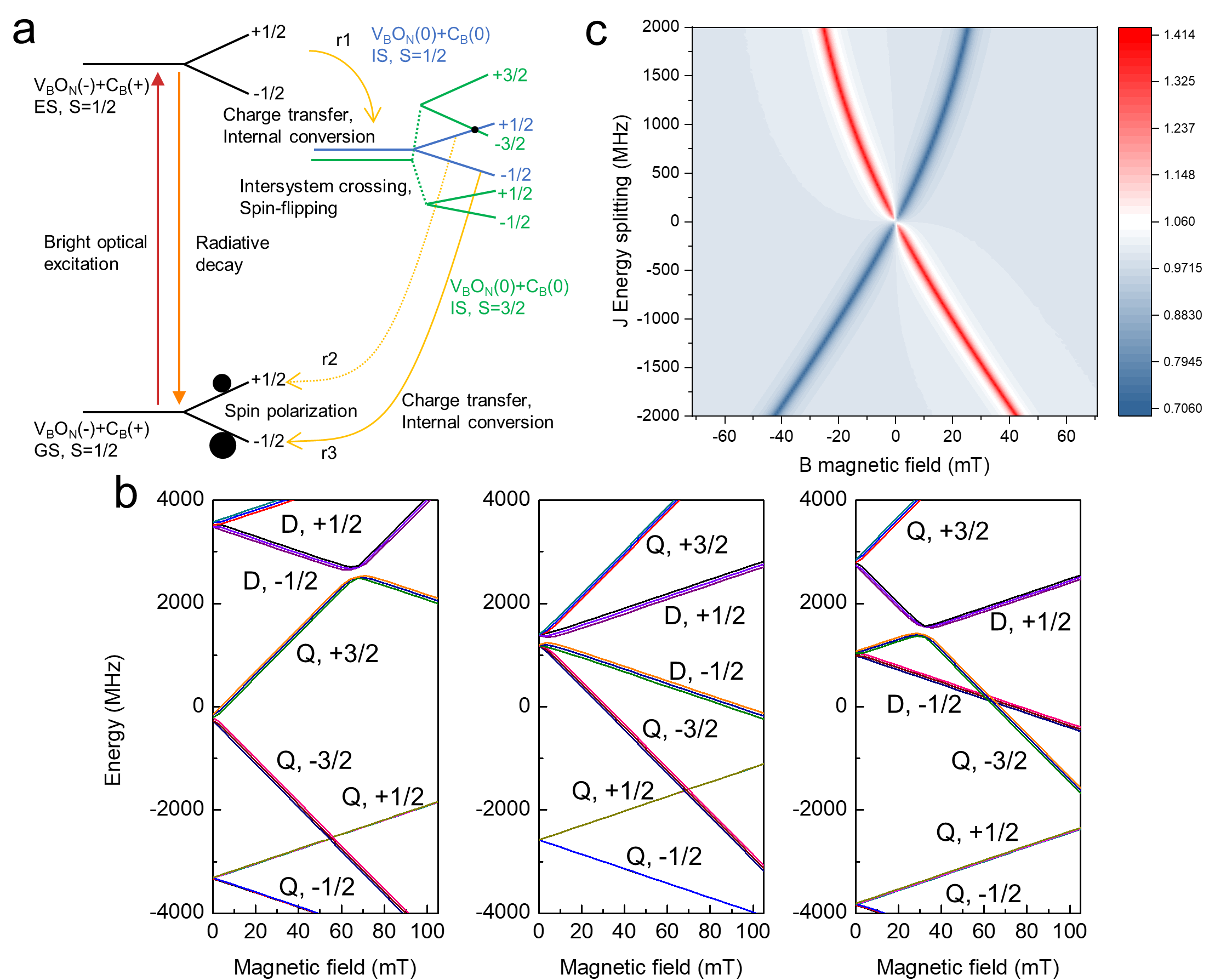}
     \caption{\label{Figure4}%
     The optical pumping loop of \cb -- \onvb. \textbf{a} Bright excitation (red) from ground state (GS) to excited state (ES) is localized on \onvb($-$). Through internal conversion, the ES state can go to doublet metastable state of \cb$^0$ -- \onvb$^0$, i.e., $D_s$. The $D_s$ are separated by $J > 0$ energy from the quartet metastable state of \cb$^0$ -- \onvb$^0$, i.e., $Q_s$. The $\pm$ 1/2 and 3/2 spin sublevels are separated by the zero-field-splitting in $Q_s$. The black dot indicates spin mixing between $D_s$ and $Q_s$. The black circle indicates the population magnitude in the ground state. \textbf{b} The spin sublevels under magnetic field in $D_s$ and $Q_s$ with $J$ values at $-3000$, $0$ and $+3000$~MHz, accordingly. \textbf{c} The population ratio between $|D_s,+1/2\rangle$ and $|D_s,-1/2\rangle$ in the metastable state at the given magnetic fields.}
\end{figure*}

The fluorescence mechanisms within DAP as shown in Figs.~\ref{Figure3}\textbf{b,c} can be classified in two types according to the relative energy levels of the donor and the acceptor. If the donor is lower than the acceptor level then the charge transfer cannot spontaneously occur. Charge transfer can be induced by illumination from $D^0$ to $A^0$ (neutral charge state) with photon energy $h\nu_a$, and this creates ionized donor $D^+$ and acceptor $A^-$. In other words, the electron from occupied state of $D^0$ is excited to the empty state of $A^0$. The electron goes back through ensuing radiative decay (charge recombination) and one photon is emitted with energy $h\nu_e$. If the donor level is higher than the acceptor level then the electron from donor may spontaneously transfer to the acceptor when they are located in short distance. Then the photoexcitation converts the charged DAP to neutral. The charge transfer from donor to acceptor is distance-dependent and a recent study investigated the charge transfer dynamics based on Marcus theory framework~\cite{lee2019donor}. Unlike the above fluorescence mechanisms within DAP, here the photoexcitation is generally active on the orbitals localized on the acceptor, which could lead to a bright emission from the excited state. In particular, this is the case for \on$^+$ -- \onvb$^-$. The shallow donor makes the levels of the dim \on$^0$ -- \onvb$^0$ states to lie at much higher energies than the bright excited state's level. In \cb$^+$ -- \onvb$^-$ DAP the situation is different, as its optical loop depicted in Fig.~\ref{Figure4}\textbf{a}. The optical excitation first drives the doublet ground state to the optically active doublet excited state. The small energy difference between the optically active excited state and the intrinsically dim metastable states enables the charge transfer from \onvb\ to \cb\ through non-radiative decay. The doublet ($D_s$) metastable level and the quartet ($Q_s$) metastable level are separated by the spin-spin exchange interaction $J$. The ground state and the optically active doublet excited states are linked to the $D_s$ state by either a weak optical transition or internal conversion (IC) while they are linked to the $Q_s$ by intersystem crossing (ISC) where spin-flip mediates the latter. By taking the same orbitals in the $D_s$ and $Q_s$ it is a common wisdom that IC is a much faster process than ISC. Therefore, the $D_s$ state will be populated in the metastable manifolds via IC. At zero magnetic field the $D_s$ state provides radiative and non-radiative pathways which shorten the optical lifetime of the \onvb$^-$ in this DAP when compared to the optical lifetime of \on$^+$ -- \onvb$^-$ DAP. We note that depending on the magnitude of the non-radiative rate ($r1$ rate, see also Supplementary Figure 6) the non-radiative transition towards the metastable $D_s$ may dominate in the decay process and then the dim optical transition may occur from $D_s$ DAP state.   


At non-zero magnetic fields the Kramers-doublets split in \cb$^+$ -- \onvb$^-$ DAP where the presence of $Q_s$ plays an important role in the spin-dependent non-radiative decay. For instance, the higher branch $m_S=+1/2$ in $D_s$ can mix with lower branch of $m_S=-3/2$ in $Q_s$. In other words, the $D_s$ acquires some $Q_s$ character and vice versa. Therefore, the $m_S=-1/2$ spin sublevel localized on \onvb$^-$ in $D_s$ can have larger population than $m_S=+1/2$ through the non-radiative decay $r3$ (see Fig.~\ref{Figure4}). This will populate the $m_S=-1/2$ spin state over the $m_S=+1/2$ spin state in the electronic ground state which makes the $m_S=-1/2$ spin level brighter than that of $m_S=+1/2$. As a consequence, ODMR signal can be observed for an $S=1/2$ ground state. The key characteristic here is the quasi-degenerate $D_s$ and $Q_s$ states in the metastable \cb$^0$ -- \onvb$^0$ configuration which is a spin pair of $S=1/2$ (see Ref.~\cite{auburger2021towards}) and $S=1$ defects (see Ref.~\cite{li2022identification}), respectively. This can lead to spin-polarization of a split $S=1/2$ Kramers ground state doublet as observed in ODMR. 

Actually, such effects for DAP systems have been already reported in wide band gap materials such as zinc-sulfide~\cite{Nicholls1979} and diamond~\cite{Nazare1995} but only phenomenological models have been provided to understand this effect. The intersystem crossing processes for spin pairs have been mostly discussed for organic chromophores~\cite{ishii1998general, kandrashkin2003light, kandrashkin2004electron, teki2020excited, qiu2023optical} which includes dipolar mixing ($\hat{H}_\text{di}$), spin-orbit-coupling ($\hat{H}_\text{soc}$) and hyperfine couplings ($\hat{H}_\text{hf}$). However, it should be noted that none of these can be easily determined due to the distance-dependence and $D_s$-$Q_s$ splitting-dependence. The $D_s$-$Q_s$ splitting scales with distance as $R^{-6}$. Relatively, the intersystem crossing caused by hyperfine interaction decays much slower than the dipolar mixing and spin-orbit-coupling.
For specified distance $R_i$, the rate to flip can be evaluated by Fermi Golden rule as
\begin{equation}
\Gamma = \frac{2\pi}{\hbar} |\langle\psi(D_s)|\hat{H}_\text{di}+\hat{H}_\text{soc}+\hat{H}_\text{hf}|\psi(Q_s)\rangle|^2 L(J) \text{,}
\end{equation}
where $\psi(D_s)$ and $\psi(Q_s)$ are the doublet and quartet wavefunction, and $L(J)$ is the lineshape function associated with the phonon overlaps between the two states. With the strongest hyperfine coupling of 55~MHz from the two $^{14}$N nuclear spins of \onvb\ in \cb5, the estimated rate has upper bound at $10^8$~Hz when we assume the doublet and quartet have nearly identical energy and geometry. This upper bound might be suppressed, e.g., via $L(J)$. Nevertheless, it still represents a sufficiently fast process for an efficient spin mixing. 

The distance between the DAP constituents should be long enough to reduce the energy gap $J$ between $D_s$ and $Q_s$ but not too long to minimize the spin-mixing rate. Unfortunately, the direct calculation of $D_s$ and $J$ is beyond the scope of the present study because it requires multi-reference methods that cannot be applied to the supercell size required for the DAP model. Nevertheless, an effective Hamiltonian for \onvb$^0$ -- \cb$^0$ with $S_1=1$ and $S_2=1/2$ coupled spin pairs at a given magnetic field $B$ may be written as
\begin{equation}
    \label{eq:spinH}
    H_\text{spin} = S_1 \mathbf{D} S_1 + \sum_i H^{(i)}_\text{hf} + g B S_1 + g B S_2 + J S_1 S_2 \text{,}
\end{equation}
where the gyromagnetic factor of the electrons is $g\approx2.0023$ (very low spin-orbit interaction) for each constituent, $J$ is the isotropic exchange interaction between the spin pairs, $\mathbf{D}$ is the $D$-tensor of \onvb$^0$ [$D=3.8$~GHz and $E=0.091$~GHz], and we only use the largest hyperfine couplings of the two $^{14}$N $I=1$ spins. We numerically solve Eq.~\eqref{eq:spinH} as a function of $B$-field (set perpendicular to the hBN sheet) and the $J$ spin exchange interaction as implemented in Easyspin~\cite{stoll2006easyspin}. 

Fig.~\ref{Figure4}\textbf{b} shows the splitting of spin sublevels under magnetic field for various $J$ values, which can exhibit different magnitude and sign depending on the relative orientation of coupled spins as appears in our DFT calculation. Due to the zero-field splitting of the triplet sublevels of \onvb\, the $|Q_s,\pm 1/2\rangle$ always lie low and the main spin mixing is from $|Q_s,\pm 3/2\rangle$ and $|D_s,\pm 1/2\rangle$. The mixing is largely due to $|D_s,-1/2\rangle$ and $|Q_s,+3/2\rangle$ when $J<0$, while the mixing is between $|D_s,+1/2\rangle$ and $|Q_s,-3/2\rangle$ when $J>0$. The different coefficient in the eigenstates of the spin sublevels implies population difference of the spin states which leads to ODMR contrast that could be tuned through constant magnetic fields. More specifically, the variation in mixing with the quartet states can result in different non-radiative relaxation rates, leading to hyperpolarization and ODMR contrast. For each finite magnitude of J-coupling, we observe a strong dependence of the ratio of mixing coefficients on the magnetic field, as illustrated in Fig.~\ref{Figure4}\textbf{b}. This difference is particularly pronounced when the applied field is set close to the avoided crossing, reaching a maximum value of approximately 1.4. Ultimately, this should translate into differences in the internal conversion rates to the ground state. It is important to note that for $J=0$, i.e., when two non-interacting defects are present, both doublet states mix equally with the quartet manifold at each magnetic field, resulting in no discernible difference in non-radiative decay.

To further substantiate the role of spin mixing in mediating the ODMR signal, we calculated the internal conversion rates towards and from the MS metastable state. We approach this process by using the excited state and MS $D_s$ for $r1$, MS $D_s$ and doublet ground state for $r2$, separately. With short distance below 1~nm, the small energy difference between the optically active doublet excited state and the dim metastable state and the strong electron-phonon coupling lead to fast internal conversion rate around $10^{10}$~MHz. Therefore the system relaxes back to ground state rapidly without detectable optical signal. The orbital overlap integral and the resulting electron-phonon coupling decay with increasing distance where the $R$ distant dependent value of the integrals may be approximated by the overlap of two Slater orbitals which goes with $e^{-R}$~\cite{silver1968overlap}. By the two configurations \cb5 and \cb7, we can extrapolate the electron-phonon coupling and conversion rate to longer $R$ distance between the donor and acceptor that cannot be directly computed with DFT, as discussed in Supplementary Note 2. The estimated conversion rate $r1$ is at 6-8~MHz with 18~\AA\ separation which results in competitive radiative and non-radiative pathways from the optically active excited state of the \cb -- \onvb\ DAP defects. This means that the ZPL wavelength, quantum yield and brightness of these DAP defects depend on the actual distance of the acceptor and donor. It is intriguing that both weak radiative and non-radiative ($r3$ internal conversion) decay pathways link the metastable state to the doublet ground state (Supplementary Note 2) that closes the optical spin-polarization loop of these DAP defects. 

With around 2~nm separation between \cb\ and \onvb, the present coupled spin pair model could explain the previously reported signature of ODMR signals from ground state $S = 1/2$ defects in hBN sheets and nanotubes~\cite{scholten2024multi, gao2023nanotube, guo2023coherent}. We underline that if the metastable state is sufficiently long-living to carry out spin operations (e.g., ST1 defect in diamond~\cite{Lee2013}) then this spin-pair model can be extended to the ODMR centers in hBN where the spin-polarization is observed in the metastable state. In this case, the ground state could be singlet from an $S = 1/2$ acceptor and $S = 1/2$ donor whereas the metastable states comprise single and triplet spin pairs. We note recent works (Refs.~\cite{patel2024, scholten2024multi,robertson2024universal}) during the preparation of our manuscript which observe this phenomenon without adding microscopic models for the underlying processes.

In summary, we propose that the DAP emission is responsible for most of the visible ZPL emission at around 2.0~eV. The variance in the ZPL wavelengths and optical lifetimes reported in the experiments can be explained by the type and the distance of the donor defect with respect to a key acceptor defect, \onvb. Our simulations reveal the ODMR mechanism associated with the spin pairs in certain DAP structures activated within the dim metastable states which is manifested as ODMR detection of the $S = 1/2$ ground state at non-zero constant magnetic fields. Our model explains the challenge to produce indistinguishable SPEs in hBN as the location of the defects pairs should be well engineered. DAPs in wide band gap semiconductors and materials are quite common structures and are responsible for optical emissions and ODMR signal for Kramers doublet spin systems. The general mechanisms proposed here is insightful to not only the defects engineering in hBN for quantum information processing but also the optoelectronic applications of defects in other semiconductors.

\section{Methods} 
\label{sec:methods}

In this paper, we carry out the density functional theory (DFT) calculations by the \textit{Vienna ab initio simulation package} (VASP) code~\cite{kresse1996efficiency, kresse1996efficient} with plane wave basis set. We applied a plane wave cutoff energy at 450~eV. The valence electrons and the core part are described with projector augmented wave (PAW) potentials~\cite{blochl1994projector, kresse1999ultrasoft}. A $8\times8$ two-layered supercell model was used to avoid the interactions between periodic images and this is sufficient to use the $\Gamma$-point sampling scheme. The interlayer vdW interaction was described with DFT-D3 method of Grimme~\cite{grimme2010consistent}. With mixing parameter $\alpha = 0.32$, the hybrid density functional of Heyd, Scuseria, and Ernzerhof (HSE)~\cite{heyd2003hybrid} could yield experimental optical gap around 6~eV without electron-phonon renormalization energy included and this functional is used to optimize the geometry and calculate electronic structures. The convergence threshold for the forces was set to 0.01~eV/\AA. $\Delta$SCF method~\cite{gali2009theory} was used to calculate electronic excited states. The band alignment in the charge correction procedure is based on the core level energy of one selected atom far away from the defect center in the defective and the perfect supercell. The zero-field-splitting due to dipolar electron-spin electron-spin interaction was calculated within the PAW formalism~\cite{bodrog2014} as implemented in VASP by Martijn Marsman. We applied the spin decontemination method to arrive at the final result~\cite{biktagirov2020}. 

\section*{Author contribution}
S.~L. and A.~G. designed the project. All authors discussed the results and contributed to the manuscript writing.

\section*{Competing interests}
The authors declare that there are no competing interests.

\section*{Data Availability}
The data that support the findings of this study are available from the corresponding author upon reasonable request.

%
%
\begin{acknowledgements}
Support by the Ministry of Culture and Innovation and the National Research, Development and Innovation Office within the Quantum Information National Laboratory of Hungary (Grant No.\ 2022-2.1.1-NL-2022-00004) as well as the European Commission for the projects QuMicro (Grant No.\ 101046911) and SPINUS (Grant No.\ 101135699) are much appreciated. AG acknowledges the high-performance computational resources provided by KIF\"U (Governmental Agency for IT Development of Hungary). 
\end{acknowledgements}

\bibliography{main}

\end{document}